\documentclass[12pt]{iopart}
\usepackage{iopams} 
\usepackage{graphicx}
\usepackage{helvet}
\usepackage{hyperref}
\usepackage{caption}
\usepackage{isotope}
\usepackage{xfrac}
\usepackage{cite}
\usepackage{color}
\usepackage{tikz}
\usetikzlibrary{arrows.meta}
\tikzset{%
  >={Latex[width=2mm,length=2mm]},
            base/.style = {rectangle, rounded corners, draw=black,
                           minimum width=4cm, minimum height=1cm,
                           text centered, font=\sffamily},
  neutron/.style = {base, fill=blue!80},
       beta/.style = {base, fill=blue!20},
    stable/.style = {base, fill=black!40},
         alfa/.style = {base,  fill=yellow,
                           font=\ttfamily},
 }                          

\begin{document}

\title{From nuclei to neutron stars: simple binding energy computer modelling in the classroom (Part 1)}

\author{ A Pastore }
\address{Department of Physics, University of York, Heslington, York, Y010 5DD, United Kingdom }
\ead{alessandro.pastore@york.ac.uk}

\author{ A M Romero}
\address{Department of Physics and Astronomy, The University of North Carolina at Chapel Hill, Chapel Hill, North Carolina 27599, USA}

\author{ C Diget }
\address{Department of Physics, University of York, Heslington, York, Y010 5DD, United Kingdom }

\author{ A Rios }
\address{Department of Physics, Faculty of Engineering and Physical Sciences, University of Surrey, Guildford, Surrey GU2 7XH, United Kingdom}

\author{ K Leech }
\address{Department of Physics, University of York, Heslington, York, Y010 5DD, United Kingdom }

\author{ P Stokoe }
\address{Department of Physics, University of York, Heslington, York, Y010 5DD, United Kingdom }

\begin{abstract}
We present a simple activity based on the liquid-drop model which allows secondary school students to explore the uses of mathematical models and gain an intuitive understanding of the concept of binding energy, and in particular the significance of positive binding energy. 
Using spreadsheets provided as Supplementary Material, students can perform simple manipulations on the different coefficients of the model to understand the role of each of its five terms. Students can use the spreadsheets to determine model parameters by optimising the agreement with real atomic mass data. 
This activity can be used as the starting point of a discussion about theoretical models, their validation  when it comes to describing experimental data and their predictive power towards unexplored regimes.
\end{abstract}

\section{Introduction}

The  \emph{Binding Blocks}  (BB) project is an educational outreach activity developed at the University of York aimed at promoting the knowledge of nuclear science to a large audience and in particular to young students\footnote{\url{https://www.york.ac.uk/physics/public-and-schools/secondary/binding-blocks/}}.
In Ref.~\cite{BB17}, some of us discussed the main outlines of the project. By using different towers of LEGO\textsuperscript{\textregistered} bricks\footnote{LEGO\textsuperscript{\textregistered} is a trademark of the LEGO Group of companies which does not sponsor, authorise or endorse the present work.}, we create a chart of nuclides in three dimensions. The towers are colour-coded according to the radioactive mechanism through which  isotopes decay. The height of each tower represents the mass excess per nucleon per kilogram of each isotope relative to $^{56}$Fe, the most stable nucleus. An online version of the nuclear chart has also been created and can be found in Ref.~\cite{Simpson2017}.

Using the three dimensional chart, we have also developed a series of activities for A-level students~\cite{Wright2017}. The primary goal is to use the striking visual impact of the chart as a formative tool for teachers to explain complicated aspects related to nuclear science. On a devoted YouTube\textsuperscript{\textregistered} channel, we have also published mini-lectures focusing on nuclear physics using this three-dimensional LEGO\textsuperscript{\textregistered} chart \footnote{\url{https://www.youtube.com/channel/UCvIXlFgJyGh4Jle\_4\_KE2aA}}.
These efforts have focused on experimental nuclear data~\cite{BB17} or on nuclear processes~\cite{frost2017cycling}, but have not addressed specifically theoretical nuclear physics ideas. 

The goal of this paper is to introduce a new series of activities that deal with the concept of a \emph{theoretical nuclear model} and its applications. We use, as a representative case, a set of relatively simple mathematical equations based on the liquid-drop (LD) model. The LD model, first introduced by George Gamow~\cite{gamow1930mass}, describes the nucleus as a drop of incompressible fluid of very high density, held together by the nuclear force. 

Our aim is to demonstrate that models can be used to describe physical data; that models often  involve a parameter optimisation process; that they can be exploited to reach conclusions far beyond their initial remit; and that they bring in their own set of (systematic) uncertainties. Theoretical model uncertainties are in fact of a different nature than those addressed in experiments, and can be explored with our proposed activities. In doing so, we expect to trigger a discussion about the fundamentals of scientific models and their applicability \cite{hickman1986mathematical,Gilbert2004}. 

\medskip

In the present activity, we specifically focus on a single nuclear physics concept - that of \emph{binding energy} (BE). Binding energies are crucial to understand why a nucleus exists, and they can be used to explain why only specific combinations of protons and neutrons are found in nature or in experimental nuclear physics facilities. BEs are routinely measured in nuclear experiments, so there is a wealth of data to compare to. Using a very intuitive theoretical model based on the LD binding energy formula, students can gain insight into the LD model, its merits and applicability. We provide a series of visualisation tools by means of Microsoft Excel\textsuperscript{\textregistered} and OpenOffice worksheets. By removing the complexity of the simulation and providing an immediate visual output, students can explore different aspects without mathematical complications. They can, for instance, use the model to explain the trends observed in nuclear data.  Importantly, they can apply the model to predict binding energies beyond those that have been measured, thus providing predictions into experimentally unknown regimes. In a follow-up contribution (Part 2), we explore how the model can be used to extrapolate beyond its initial remit by discussing neutron stars \cite{Part2_2020}.  


Across the paper, we propose a series of questions and challenges that can be addressed by students. Some of these do not require access to the computational worksheets, and can therefore be addressed directly in a classroom presentation. The worksheet activities require access to a computer, but allow for an interactive, and more active, learning experience. We provide two different worksheets that can be used in different settings. \verb+Worksheet_Instructor+ contains data for several isotopic chains, and can be run in in classrooms and computational laboratories, or as outreach activities in scientific centers with the guidance of teachers, instructors or tutors. \verb+Worksheet_Student+ is designed to be a standalone online activity, that individual students can run in an online setting - be it at home or in a classroom. This can be accompanied by a minimal instruction set and introduced by online means (e.g. a video like the one here\footnote{https://www.youtube.com/watch?v=Qsu7IrGiOIk}). 

The article is organised as follows. In Sec.~\ref{sec:theo}, we introduce the concept of binding energy, while in Sec.~\ref{sec:LD} we present the LD model equations. Section~\ref{sec:opt} describes simple paramater optimisation strategies for the model. In Sec.~\ref{sec:limit} we illustrate how to predict the binding energies of not yet measured nuclei. We present our conclusions in Sec.~\ref{sec:concl}. 


\section{Binding energy}\label{sec:theo}

The binding energy of a physical system is the amount of energy required to separate any composite system into all of its constituents. 
The concept of binding energy is ubiquitous in science, and is particularly relevant for understanding the chemical properties of atoms and molecules; the astrophysics of gravitational systems and the physics of nuclei. 
In the case of a nucleus, the constituents are the individual $Z$ protons and $N$ neutrons, inclusively called \emph{nucleons}. The mass number of the nucleus is $A=N+Z$. The binding energy of an isotope of element $Z$ with $N$ neutrons is defined by the difference
\begin{eqnarray}\label{eq:BE}
BE&=N m_n c^2 + Z m_p c^2 - M_{N,Z} c^2 \, ,
\end{eqnarray}
where $m_nc^2= 939.565$ MeV\footnote{In nuclear physics activities, it is more natural to work in units of mega-electronvolts (MeV)  rather than Joules (J). We recall that  $1$ J=$6.242\cdot 10^{12}$ MeV} and $m_pc^2=938.272$ MeV represent the rest mass energies of the neutron and the proton, and $M_{N,Z}$ is the rest mass of the isotope itself. 
Reference~\cite{baroni2018teaching} provides a more detailed discussion and some simple practical activities to familiarise interested readers with the concept of binding. 
It is important to differentiated binding energies from separation energies, which instead correspond to the minimum energy required to remove one (or more, but not \emph{all}) the constituents of the composite system. 
The typical binding energy per particle of a medium-to-heavy nucleus is of the order of $\approx 8$ MeV.

Using advanced experimental techniques~\cite{Lunney2003,kluge2004trapping,mukherjee2008isoltrap}, it is possible to measure the binding energies of very short-lived radioactive nuclei, far away from the valley of stability. 
According to  the most recent nuclear database~\cite{wang2017ame2016},  more than $2400$ nuclear binding energies per particle have been measured with an accuracy of more than $0.005\%$. With such remarkable levels of accuracy, the error bars would not be visible on the figures illustrating experimental values. We have thus decided not to consider the presence of experimental errors in the activities described below.


\begin{figure*}[!t]
\begin{center}
\includegraphics[width=0.6\textwidth]{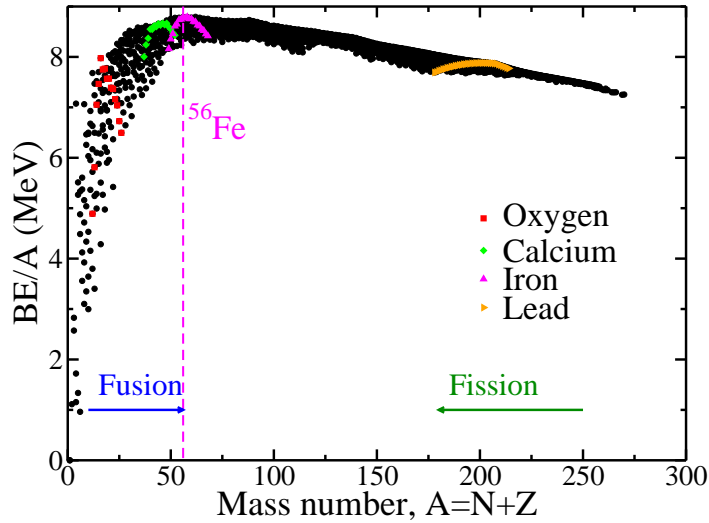}
\end{center}
\caption{(Color online) Binding energy per nucleon, $BE/A$ in MeV, as a function of mass number $A$. Data points come from the experimental measurements compiled in Ref.~\cite{wang2012}. See text for details.}
\label{fig:be}
\end{figure*}

We illustrate the evolution of the experimental binding energy per particle, $BE/A$, as a function of mass number $A$ in Fig.~\ref{fig:be}. The black circles correspond to the data for $2411$ isotopes from Ref.~\cite{wang2012}. 
On the same plot, we highlight some particular isotopic chains: O ($Z=8$), Ca ($Z=20$), Fe ($Z=26$) and Pb ($Z=82$).
A number of interesting conclusions can be drawn from this figure alone, which could very well be the starting point for the discussion of the activities we propose below. 
When illustrating this figure to students, demonstrators should highlight that $BE/A$ is not a flat line as a function of $A$, but rather a curve 
with  a maximum in the region $A=56-62$. We illustrate this peak with the more abundant isotope of iron, $^{56}$Fe, rather than rarer (but most bound overall) nuclide $^{62}$Ni \cite{Shurtleff1989}. 
By moving from the origin up to $^{56}$Fe, the $BE/A$ plot shows that if two light nuclei fuse together, they can gain binding energy~\cite{frost2017cycling}. Exothermic fusion processes are relevant for fusion energy and for stellar astrophysics, where  they prevent the gravitational collapse of stars \cite{ostlie2006introduction}. 
In contrast, beyond $^{56}$Fe, the fusion of two nuclei is an endothermic process. This explains why nuclei in the mass region $A=56-62$, like $^{56}$Fe, are found as end-points in the fusion cycle of a star, and are copiously produced in type Ia supernovae \cite{Shurtleff1989,ostlie2006introduction}.
For very massive nuclei, as Uranium for example, the opposite is true. By splitting an isotope into two smaller nuclei, the binding energy of the two  fragments is larger than the one of the parent nucleus~\cite{hodgson1997introductory}. This well-known mechanism is called nuclear fission, and it is harnessed in nuclear reactors to produce energy~\cite{stacey2018nuclear}. For a more detailed discussion on these topics, we refer the reader to Ref.~\cite{Wright2017}.

In addition to the ideas associated to fusion and fission, instructors can use the data of Fig.~\ref{fig:be} to address other relevant questions, like:
\begin{itemize}
\item How many nuclei exist in total? 
\item What is the maximum number of protons we can find in a nucleus?
\item What is the heaviest isotope we can find for each chemical element?
\end{itemize}
The interest of these  questions is not purely academic. At present, about $\approx 3400$ isotopes have been discovered \cite{Nubase2016}, but the number of new isotopes increases by about $\approx 10-30$ per year \cite{Thoennesen_book,Thoennessen2011}. Some of these isotopes are relevant for a variety of purposes, including nuclear medicine~\cite{mettler2011essentials}. The teacher can refer also to the recent discovery of the new chemical elements $Z=117$~\cite{oganessian2010synthesis} and $Z=118$~\cite{Oganessian2006}. The latter was recently named Oganesson by the IUPAC, in honour to the superheavy-element scientist Y. Oganessian, who discovered it \cite{generaliupac,oganessian2007heaviest}.

Knowledge on the limits of the chart of atomic nuclei is also relevant to current astrophysical research. In particular, the rapid neutron capture process (or r-process) is partly responsible for the nucleosynthesis of nuclei above $^{56}$Fe~\cite{koura2014three}. This process takes place in the neutron-rich side of the nuclear chart, way beyond the stability valley, and in some cases also beyond the limits of presently known isotopes. The very recent discovery of gravitational waves from a neutron star merger~\cite{abbott2017multi} and its counterpart electromagnetic signal has unambiguously proven that these mergers are sites for the r-process~\cite{Watson2019} - something that was only predicted by theory before~\cite{goriely2011r}. 

All in all, there is a strong motivation to perform experiments that can extend our knowledge of the nuclear landscape. Measuring new nuclei is, however, not an easy task. Moving away from the valley of stability, the lifetime of nuclei becomes shorter and shorter. This requires both innovative production methods as well as complex detection techniques. It is therefore important to have theoretical guidance indicating which nuclei are bound, and can be detected. In addition, to address important astrophysical questions, it is crucial to develop mathematical models~\cite{hickman1986mathematical} that can reliably reproduce current data, but can also be used to predict new phenomena. Prediction and validation are key steps of the scientific process. Ultimately, if models are robust enough, one can use them to predict the properties of nuclei in regions of the Segr\'e chart where no experimental evidence exists.

\section{The Liquid Drop model}\label{sec:LD}

\begin{table}[!t]
\begin{center}
\begin{tabular}{|c|c|c|}
\hline
\hline
 \multicolumn{2}{|c|}{Coefficients [MeV]}\\
\hline
$a_V$& 15.8   \\
$a_S$&   18.3  \\
$a_C$& 0.714  \\
$a_{A}$& 23.2   \\
$a_{P}$& 12.0  \\
\hline
\hline
\end{tabular}
\end{center}
\caption{Coefficients (expressed in MeV) of the LD mass formula Eq.~(\ref{eq:LD}). }
\label{tab:coeff}
\end{table}

Within the scientific literature, it is possible to find a variety of nuclear models that have been developed to analyse nuclear BEs~\cite{sobiczewski2014accuracy}. By ``model", here, we specifically refer to a mathematical formula that is optimised to fit a restricted dataset, and can be used to provide an understanding of data and to extrapolate to unknown domains. We provide relatively simple worksheets that can help instructors develop useful teaching tools which address different aspects of the scientific practice. The expected learning outcomes of these worksheets are:
\begin{itemize}
\item To familiarise students with the concept of theoretical mathematical models. 
\item To introduce the concept of stability and the nuclear landscape.
\item To develop simple computational skills to perform predictions based upon numerical applications of the model.
\end{itemize}
To this purpose, we have select  a simple equation based on the LD model and used to obtain the well known Bethe-Weizs\"acker (BW) mass formula~\cite{hodgson1997introductory}. Before entering into a detailed discussion of the model, it is worth recalling that \emph{any} model is based on physical insight of the system under study. In the case of the LD discussed here, one assumes that protons and neutrons form an incompressible droplet with a relatively sharp surface. The BE is only function of neutron number $N$ and proton number $Z$, which appear individually in the formula or added up in the mass number variable $A$. In particular, the model yields the binding energy per particle of an atomic nucleus as the sum of five terms:
\begin{eqnarray}\label{eq:LD}
\frac{BE}{A}&=a_V-\frac{a_S}{A^{1/3}}-a_C\frac{Z^2}{A^{4/3}}-a_{A}\frac{(N-Z)^2}{A^2}+a_P\frac{\delta_{N,Z}}{A} \, ,
\end{eqnarray}
\noindent where
\begin{eqnarray}
\delta_{N,Z}= \left\{ \begin{array}{cc}
 \frac{1}{A^{1/2}}, & \text{N is even and Z is even}\\
  -\frac{1}{A^{1/2}}, & \text{N is odd and Z is odd}\\
  0, &  \text{N is even and Z is odd}\\
    0, &  \text{N is odd and Z is even}\\
\end{array}\right.
\end{eqnarray}
The multiplicative coefficients in front of each term, $a_X$, are provided in Tab.~\ref{tab:coeff} and taken from Ref.~\cite{Weiz1935}. We refer to Ref.~\cite{pastore2019introduction} for a detailed discussion on how these parameters are usually determined.

\begin{figure*}[!t]
\begin{center}
\includegraphics[width=\textwidth,angle=0]{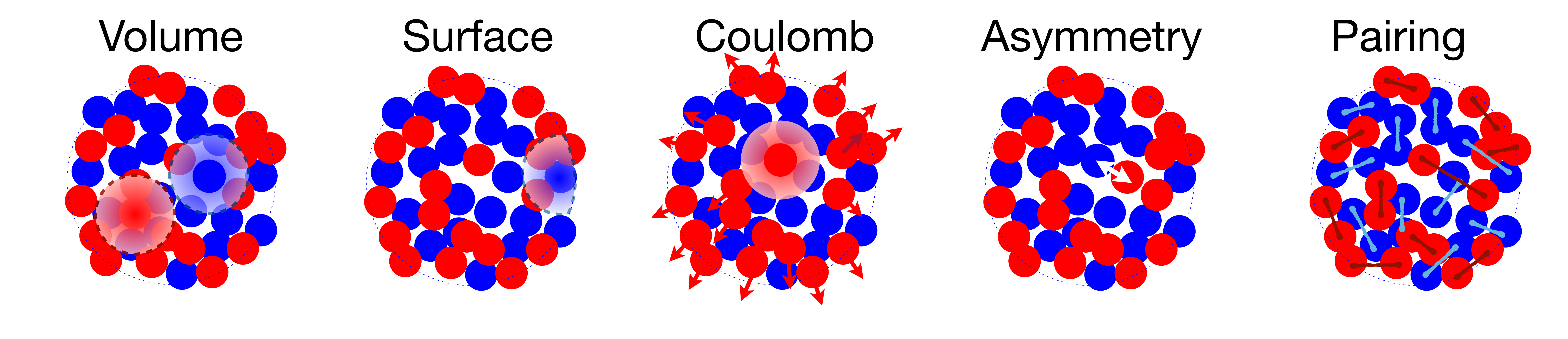}
\end{center}
\caption{(Color online) Graphical illustration of the various terms of the LD formula given in Eq.~(\ref{eq:LD}). 
}
\label{cartoon:LD}
\end{figure*}

A key advantage of using the LD mass formula is that we can associate a physical meaning to each term appearing in Eq.~(\ref{eq:LD}) \cite{Heyde_book}. The arguments are relatively simple and only require a general understanding of nuclear physics terms. A deeper understanding of the analytical forms may require the relation between nuclear radius and mass number, using the formula
\begin{eqnarray}
R=r_0 A^{1/3}\;,
\label{eq:radius}
\end{eqnarray}
\noindent 
where the parameter $r_0=1.2$ fm is obtained by fitting experimental matter radii \cite{Heyde_book,krane1988introductory,hodgson1997introductory}. In Fig.~\ref{cartoon:LD}, we provide a graphical illustration of the five terms involved in the LD formula. The physical content of each term is summarised below:
\begin{itemize}
\item Volume term (coefficient $a_V$): nucleons are subject to the nuclear strong force only if they are very close to each other~\cite{burge1967misconceptions}. Because of this short-range nature, the total binding energy of the nucleus should scale approximately linearly with the number of nucleons, with each nucleon contributing a constant, positive amount, $a_V$.
\item Surface term (coefficient $a_S$): the nucleus has a relatively sharp surface. Nucleons at the edge do not interact with as many nucleons as those in the interior. The surface term corrects the volume term to account for the smaller contribution of such surface nucleons. This term can be explained in terms of a nuclear surface tension, which grows approximately with the size of the nuclear surface, $A^{2/3}/A \approx A^{-1/3}$. 
\item Coulomb term (coefficient $a_C$): protons repel each other via the electromagnetic Coulomb interaction. The more protons in a nucleus, the larger the Coulomb repulsion and the lower the binding energy. Since the electromagnetic interaction has a long range and acts at \emph{any} distance, the energy scales as the square of the number of protons, $Z^2$. 
\item Asymmetry term (coefficient $a_{A}$): this term is related to the Pauli principle~\cite{messiah1958quantum,malgieri2016feynman}. Two fermions (in this case, nucleons) can not occupy the same quantum state.
This implies that if we increase the difference between protons and neutrons, an excess of nucleons is forced to occupy levels at higher energies, making the nucleus more unstable. This term therefore provides a penalty in the binding energy for systems that are asymmetric in isospin, $N \ne Z$.
\item Pairing term (coefficient $a_P$): nucleons in nuclei are eager to form coupled pairs, with anti-aligned spins in order to increase the binding energy~\cite{bardeen1957microscopic,bohr1958possible}. This pairing mechanism is akin to electron pairing in superconducting systems. This is represented in terms of a function $\delta_{N,Z}$ that brings in more binding when $N$ and $Z$ are even, and decreases binding when $N$ and $Z$ are odd. 
\end{itemize}
The last two terms of the LD are intrinsically related to the quantum features of the atomic nucleus. They require physical knowledge which is likely at an undergraduate level. Having said that, the formula is simple in mathematical terms, and a deep physics understanding of each individual term is not necessary to engage with the activities in the worksheet. 

\begin{figure*}[!t]
\begin{center}
\includegraphics[width=0.6\textwidth]{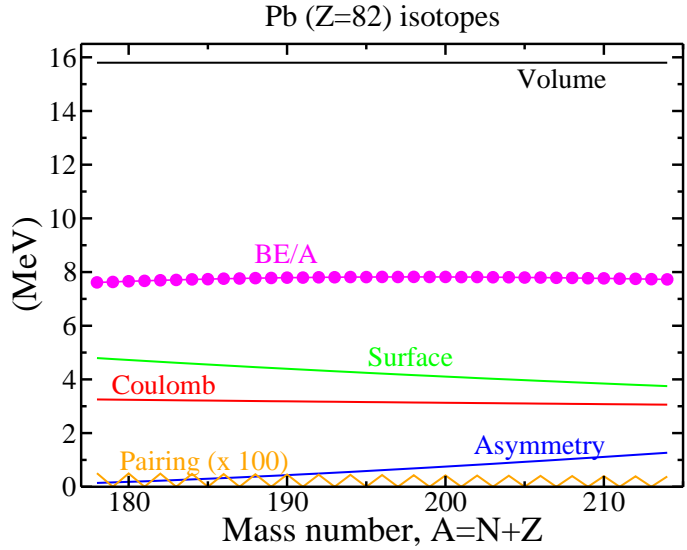}
\end{center}
\caption{(Color online) Contribution to the total binding energy per particle of the different terms appearing in Eq.~(\ref{eq:LD}) for Pb isotopes as a function of mass number, $A$. Solid circles refer to the total binding energy per nucleon. The sign in front of the contributions is neglected for illustrative purposes.} 
\label{mass2}
\end{figure*}

We illustrate the contribution to $\text{BE/A}$ of the five different terms for Pb ($Z=82$) isotopes in Fig.~\ref{mass2}  . 
The overall binding energy per nucleon (solid circles) is approximately $ 8$ MeV across the whole isotopic chain. In the scale of the graph, this appears to be relatively constant although, in fact, it has a maximum around $A \approx 198$ and changes by about $0.2$ MeV from its minimum to its maximum values. 
As discussed previously, the volume term is constant and it gives a positive contribution of approximately 16 MeV. 
In theoretical studies, one often considers a system akin to a nucleus that has an infinite extent; an equal number of neutron and protons; and the electromagnetic repulsion switched off. This artificial system is called \emph{nuclear matter} and is a useful testing ground for several theoretical considerations~\cite{baldo1999nuclear}. One typically assumes that the binding energy per particle of such system should equal $a_V$, although caveats have been raised about the extrapolation of the LD formula in the limit $A \to \infty$~\cite{atkinson2020}.

The surface and Coulomb term introduce major corrections to the binding energy, of the order of $3-5$ MeV. In contrast, the asymmetry term starts becoming relevant only when there is a a strong imbalance between the neutron and proton number at relatively large values of $A$. Even then, its contribution is well within $\approx 1$ MeV in size. The pairing term is multiplied by a factor of $100$ to make it  visible in the figure. This term therefore represents a minor correction, and it may eventually  be discarded to further simplify the model without loosing too much accuracy.


To get a better understanding of the LD model, we suggest two activities based on 
relatively simple spreadsheets, in both Microsoft Excel\textsuperscript{\textregistered}  and OpenOffice. 
\verb+Worksheet_Instructor+ contains experimental information on several isotopic chains and allows for wider applications of the formula. Because of the relatively large amount of information in this spreadsheet, we expect that students will require guidance by tutors to work with it. As such, this spreadsheet is devised as a hands-on activity with tutor support in a shared environment (like a computer classroom). In contrast, \verb+Worksheet_Student+ provides a simpler and self-contained exploration of the LD model, and has been devised as an individual, online activity. Both worksheets facilitate the following three activities:
\begin{enumerate}
\item An \emph{optimisation} exercise of some of the coefficients of the LD formula,
\item A \emph{validation} process in terms of  experimental data, and
\item A \emph{prediction} assignment, looking for the limits of nuclear stability.
\end{enumerate}
The following section discusses each one of these activities in more detail.

\section{How are liquid drop models optimised?}\label{sec:opt}

The first activity is focused around the idea of optimisation of the parameters of a model. In other words, once a mathematical model has been established, it is important to determine the coefficients (or coupling constants) of the model. 
It is worth recalling that, while the terms of the LD model are fixed by the hypothesis we made to derive the model, the coefficients ($a_V,a_S,\dots$) are adjusted to reduce the discrepancy between the theoretical predictions and the observed data.
The learning outcomes of this activity are:
\begin{itemize}
\item To introduce the concept of a mathematical model.
\item To understand the concept of parameter optimisation with a simple example.
\item To analyse the data and identify patterns and trends that are described by a mathematical model.
\end{itemize}

In Tab.~\ref{tab:coeff}, we provide the coupling constants appearing in the LD formula. These values are obtained by fitting the model over a selected pool of experimental binding energies. 
Rather than using complex statistical techniques, we want students to get acquainted with the idea of parameter optimisation by simple means, using an intuitive graphical interface. This immediate visualization is provided by figures in each tab of our spreadsheet.

\begin{figure*}[!t]
\begin{center}
\includegraphics[width=1.1\textwidth,angle=0]{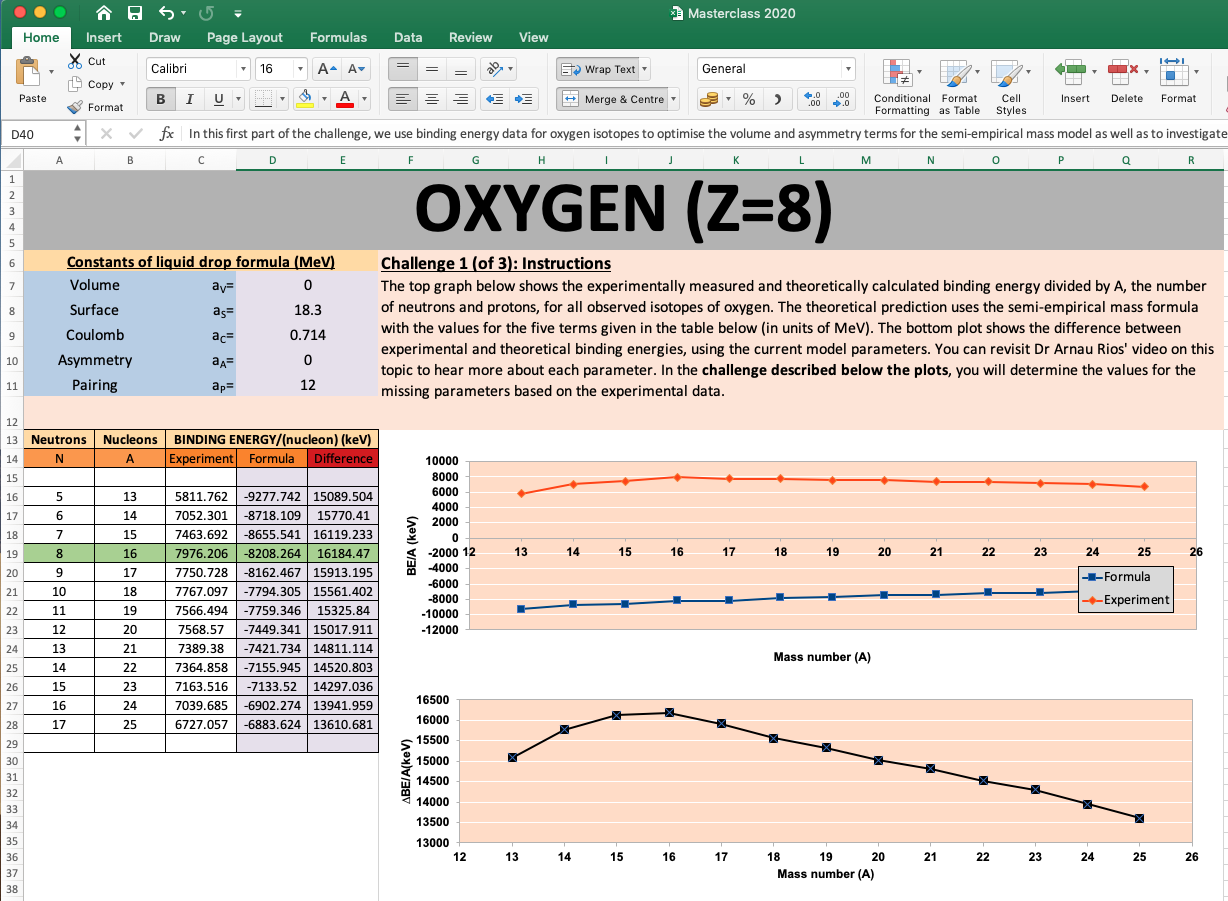}
\end{center}
\caption{(Color online) Screen shot of the  Microsoft Excel$\textsuperscript{\textregistered}$ version of \texttt{Worksheet\symbol{95}Student} used for this optimisation activity. }
\label{mass:excelA}
\end{figure*}

The first tab of the Microsoft Excel\textsuperscript{\textregistered} \verb+Worksheet_Student+ is shown in Fig.~\ref{mass:excelA}. This standalone worksheet is divided intro three different ``challenges" - each one in a different tab in the spreadsheet. Challenge 1 is used to determine two of the coefficients in the formula, $a_V$ and $a_A$, by using existing nuclear binding energy data of oxygen isotopes. The oxygen isotopic chain is rather robust for independent studies of LD coefficients, since the dependency on the volume term is not quite as strong and the limits for ``acceptable" asymmetry term values can be rather well constrained.

The two parameters $a_V$ and $a_A$ are originally set to zero in the spreadsheet, and the student is requested to find acceptable values following the protocol discussed below.
By inspecting Eq.~(\ref{eq:LD}), one notices that the asymmetry term does not contribute in $N=Z$ nuclei. As a consequence, we suggest to adjust the value of the volume term using the binding energy of a single isotope, $^{16}$O, which is highlighted in green in the table of binding energy data. This optimisation is easily performed by a trial and error procedure, following a suggested initial range of $a_V$ values. 

Once the volume term is fixed, the asymmetry coefficient $a_A$ can be adjusted to reproduce the experimental masses along the entire isotopic chain.  This can be done again by trial and error on the $a_A$ constant, until the shape of the theoretical binding energy curve matches the experimental data as well as possible.
The demonstrator (or the online material) can give a \emph{reasonable} range of values for such a parameter, typically $a_A\in[10-20]$ MeV.

Finally, students are asked to switch off the pairing term, setting $a_P=0$. The student will then observe the appearance of a zig-zag behaviour in the data - the so-called odd-even staggering of BEs \cite{Heyde_book}. The emergence of such a trend motivated the inclusion of the pairing term in the microscopic mass formula of Eq.~(\ref{eq:LD}), well in advance to the discovery of the theory of superconductivity~\cite{bardeen1957microscopic} and its application to the case of nuclear physics~\cite{bohr1958possible}.

Alternatively, the extended datasets in \verb+Worksheet_Instructor+ can be used to provide a wider analysis of nuclear data with the help of tutors or demonstrators. We envisage the use of this worksheet in a larger computer classroom activity, where students may be divided into different working groups. This file is composed of multiple tabs containing various comparisons of the LD model predictions to experimental data. The tabs named after some well known isotopic chains, namely \verb+Oxygen+ ($Z=8$), \verb+Calcium+ ($Z=20$), \verb+Nickel+ ($Z=28$), \verb+Tin+ ($Z=50$) and \verb+Lead+ ($Z=82$), are to be used in this first optimisation and validation activity. In this spreadsheet, the $5$ LD coefficients are set to their optimal values from the start. 
Students should compare the predicted binding energies with the experimental ones.
To identify and understand specific patterns in the data is a key element in the construction of \emph{any} nuclear model.

Demonstrators can ask the students to freely modify the parameters of the LD formula provided in the file, and observe how those changes affect the reproduction of experimental values. 
We note here that, by default, the constants provided in the worksheets have been obtained using a \emph{global} adjustment, \emph{i.e.} using information from all known masses. It is therefore  possible to perform local adjustments (or fits) to improve the reproduction of a specific isotopic chain. In activities with large student numbers, tutors could assign different isotopic chains in \verb+Worksheet_Instructor+ to different student groups to avoid repetition. The optimisation process could be ``gamified" by requesting each group to minimise the theory-experiment discrepancies, and establish a competition for the ``best" parameters sets. One can then compare the values obtained in different chains, as a starting point for a discussion about statistical and systematic errors in the fit. 

After students have optimised their fit using data with either worksheet, they can \emph{validate} it by comparing the predictions of the model to the experimental data.  
We recall here that the typical accuracy of the LD model is $0.1$ MeV per particle, so it should be made clear that discrepancies of this size are acceptable in the optimisation and validation procedure.

\section{Limits of nuclear stability}\label{sec:limit}

In this section, we discuss the concept of \emph{prediction} based on a mathematical model. In other words, having optimised the coefficients in a given mathematical expression, one can use the model to conjecture new ideas. 
If these predictions sit within the range of data used to fit the model, one typically states that this is an ``interpolation". In contrast, if the predictions lie outside of the optimisation data set, the model we are using is   extrapolating. For instance, using the LD model presented before to discuss the BE energy of unknown nuclei with similar masses can be considered a prediction. In contrast, the use of this model to discuss the neutron-rich dense matter inside neutron stars is clearly an extrapolation procedure. 
Extrapolation is generally less reliable than interpolation, but it is nonetheless a key aspect of the scientific method. The second paper in this series will look in more detail to these extrapolation procedures in the context of neutron stars.

The learning outcomes of this section are:
\begin{itemize}
\item To introduce the concept of prediction within a scientific model. 
\item To understand the concept of stability of a nuclear system.
\item To explain ideas around the validity of scientific models. 
\end{itemize}

\begin{figure*}[!t]
\begin{center}
\includegraphics[width=0.7\textwidth]{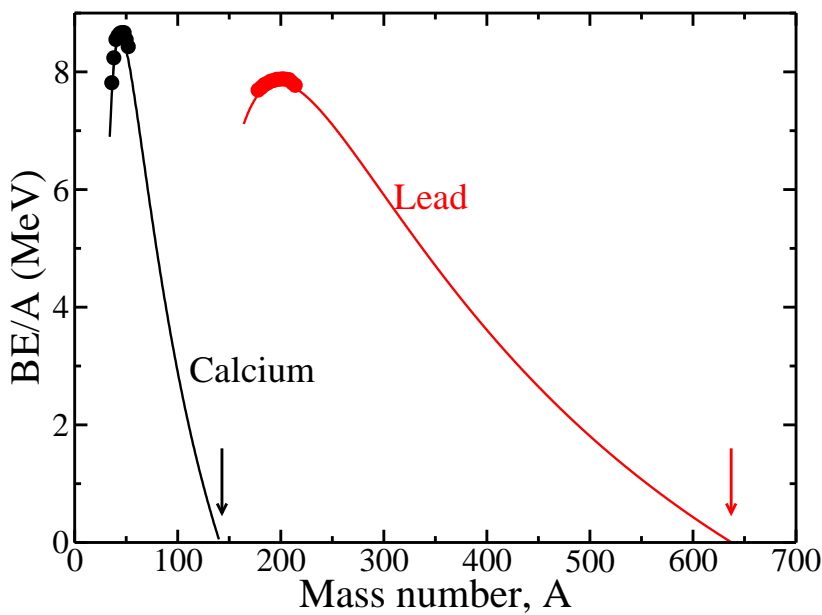}
\end{center}
\caption{(Color online) Binding energy per nucleon, $BE/A$ in MeV, as a function of mass number, $A$, for the Calcium ($Z=20$) and Lead ($Z=82$) isotopic chains. Lines represent the results obtained with the LD formula in Eq.~(\ref{eq:LD}). Symbols refer to measured experimental values.  The arrows point to the predicted maximum bound nucleus.}
\label{mass1}
\end{figure*}

Having optimised and validated the model in the previous activity, students can now turn and use it to make predictions. For a given isotopic chain (constant $Z$), one can increase the number of neutrons $N$ to systematically reduce $BE/A$. At a given maximum neutron number $N_\text{max}$ (or a corresponding maximum mass number $A_\text{max}$) the binding energy changes sign. A nuclear configuration with a negative $BE$ is unbound with respect to the decay of all its constituents.

In Fig.~\ref{mass1}, we illustrate the evolution of the binding energy per particle for two isotopic chains: Calcium ($Z=20$) and Lead ($Z=82$). We have used for this purpose the parameters of the LD model given in Tab.~\ref{tab:coeff}. On the same figure, we also provide the available experimental masses (filled circles). A very striking observation is that the number of nuclei predicted by the LD model is much larger than the number of measured masses. Demonstrators may use this figure as an introduction and may highlight the following two main features:
\begin{itemize}
\item For any chemical element, several tens (or even hundreds) of isotopes may exist.
\item Despite all experimental efforts, the number of nuclei that may exists is still much larger than the number of nuclei observed.
\end{itemize}
For calcium, the set of parameters in Tab.~\ref{tab:coeff} predicts $A_\text{max} = 140$, whereas for lead one gets the surprisingly large value of $A_\text{max}=636$. If students use their own fits for the LD coefficients derived in the previous activities, these numbers can typically change by $10-20$ mass units. This can give rise to a discussion about the systematic errors associated to predictions.

These predictions are of course not fully reliable due to the several approximations used to derive Eq.~(\ref{eq:LD}) and, importantly, due to the nature of nuclear stability itself. Typically, neutron-rich isotopes with $N \ll N_\text{max}$ are already unbound to other constituent decays. In the neutron-rich side of the table, the dominant mechanism is likely to be one-neutron emission. In other words, above a given $N_\text{drip}$, the isotopes of a given element are unstable to the decay of a neutron. 
The boundary between bound and one-neutron unbound isotopes correspond to the so-called \emph{neutron dripline}. Similarly, on the proton-rich side of the nuclide chart, there is a line that separates bound and proton unbound nuclides. This is the \emph{proton dripline}.

Neutron and proton driplines are typically used to denote the limits of nuclear stability in the Segr\`e chart. Even though typically $N_\text{drip} \ll N_\text{max}$, the neutron dripline has only been confirmed experimentally up to $Z \approx 8-9$. 
Advanced theoretical nuclear models~\cite{goriely2013hartree} predict the limit of existence of the calcium and lead isotopic chains at $^{70}$Ca and $^{274}$Pb. We note however that even these models have a relatively large systematic errors, and several current research efforts are aimed at improving the quality and reliability of such predictions~\cite{erler2012limits,Neufcourt2020}. 
For simplicity, we keep the discussion in terms of stability at the level of the change in sign of $BE/A$, to avoid having to introduce further nuclear physics ideas.

\begin{figure*}[!t]
\begin{center}
\includegraphics[width=0.8\textwidth,angle=0]{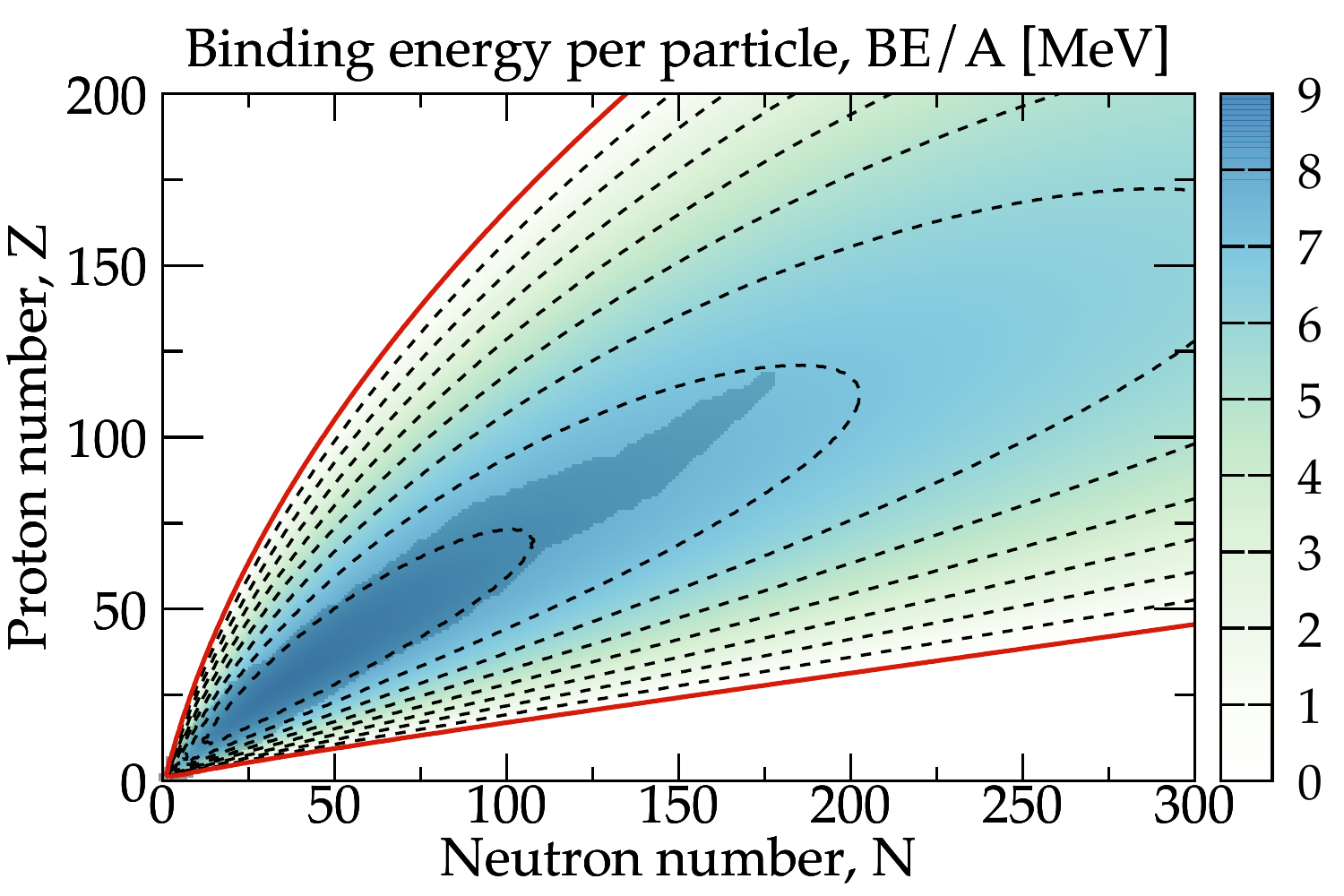}
\end{center}
\caption{(Color online) Binding energy per nucleon in MeV as a function of neutron number, $N$, and proton number, $Z$, obtained from Eq.~(\ref{eq:LD}). The dashed lines give contours in the range from $1$ to $8$ MeV in spaces of $1$ MeV. The contour at $BE/A=0$ is shown with a solid (red) line. The background black shape represents the nuclear chart of $\approx 3400$ known isotopes according to Ref.~\cite{Nubase2016}. }
\label{mass}
\end{figure*}

Keeping in mind that the LD model is a simple approximation, one can also build the entire mass table from it. In other words, we can look at the full evolution of the theoretical binding energy per particle as a function of $N$ and $Z$ from Eq.~(\ref{eq:LD}).
We present the results for this theoretical exercise in the density contour plot of Fig.~\ref{mass}. Dash-line contours are spaced by $1$ MeV in $BE/A$, and the solid line represents the limit of stability at $BE/A=0$. We also show the Segr\`e chart of experimentally detected isotopes according to the NuBASE 2016 compilation~\cite{Nubase2016}.
We observe that the LD model predicts the existence of many more nuclei than actually measured experimentally, as expected. This figure can be used by demonstrators as a starting point for several discussions, including ideas around the differences in the limits of nuclear stability in the neutron- and proton-rich sides of the chart; the total number of bound isotopes; and the shapes of the contour lines for $BE/A$.

The two worksheets we provide can be used to answer relevant nuclear physics questions in different settings. \verb+Worksheet_Instructor+ provides a more complete set of predictions for the five isotopic chains discussed earlier - in the tabs called \verb+Prediction (Oxygen)+, \verb+Prediction (Calcium)+, etc.
As such, these tabs can be used to answer questions like:
\begin{itemize}
\item Which is the heaviest isotope (fixed number of protons) of oxygen ($Z=8$)?
\item Which is the heaviest isotope (fixed number of protons) of lead ($Z=82$)?
\item Which is the heaviest nucleus with $N=Z$ we can produce with the LD formula?
\end{itemize}
All these question rely on finding the combination of $N,Z$ for which Eq.~(\ref{eq:LD}) becomes equal to zero. This procedure is implemented approximately in the spreadsheets by calculating the binding energy for several $N$ ($Z$) configurations and finding where $BE/A$ changes sign. 
Students may be asked to explore the differences between the global fit provided initially and their own (local) fits derived in the first activity.
We leave it to the demonstrators to adapt the questions and spreadsheets to the needs of the teaching. 

To expand the activity, the demonstrator may use the \verb+Prediction N=Z+ tab, which uses the LD formula to find the limit of stability for isotopes with $N=Z$. This may be used in the context of a discussion about recent discoveries of super-heavy elements \cite{oganessian2007heaviest,oganessian2010synthesis,Oganessian2006,hofmann2000discovery}. 
We point out again that these activities are easily amenable to a group setting, and may be used as the start of a competition. 
An advanced activity for interested students could involve an analytical derivation of $N_\text{max}$ or $A_\text{max}$ from Eq.~(\ref{eq:LD}), as typically discussed in undergraduate textbooks~\cite{Heyde_book}. 

In contrast, the discussion in \verb+Worksheet_Student+ is focused in a single isotopic chain, that of oxygen, as we expect that this will be easier to tackle at an individual level in an online environment. 
By using the parameters adjusted in the first activity, students should be able to predict the heaviest oxygen isotopes that may exist in nature. The accompanying text to the activity or the online demonstrators should stress here that predictions are sensitive to the choice of the parameters $a_V$ and $a_A$. To this purpose, this second challenge asks to explore the robustness of the prediction by performing small variations on the $a_V$ and $a_A$ parameters determined in the previous section. This method is usually called \emph{sensitivity} analysis, and is typically performed \emph{a posteriori} in optimisation setups. This type of analysis is used to determine the parameters that are most important for predictions \cite{Bertsch2017,pastore2019introduction}.

\section{Conclusions}\label{sec:concl}

We have presented a simple activity for young students focused on the concept of \emph{modelling}. By using a simple mathematical formula based on few physical considerations, it is possible to explain a large variety of nuclear data for binding energies. 
The activity requires the use of simple spreadsheet software.
This intuitive and interactive setting allows for a discussion on 3 key stages of theoretical modelling: optimisation, validation and prediction. Using the worksheets, students can visually optimise the model and explain observed trends in nuclear data. They can also predict the limits of nuclear stability for different isotopic chains. 

This provides an operational, and thus hopefully deeper, understanding of theoretical modelling in (nuclear) physics, which often involves these 3 stages. Importantly, the activity also provides a feel for the order of magnitude of BEs; for the different terms within the LD formula; and also for the number of bound isotopes per isotopic chain. In a follow-up paper, we use the model to extrapolate into the unknown domain of neutron-rich, macroscopically large objects called ``neutron stars". These have been of course been observed by different astrophysical means and have their own relevance. 

We provide two different spreadsheets, so that each stage can be explored either individually or in group settings. 
It is relatively easy to modify these activities, so demonstrators may want to  extend them according to the physics and computer skills of the classroom. 
At the end of the activity, demonstrators may also want to test the success of the learning experience. We are developing a series of evaluation tools (quizzes and questions) that may provide such an evaluation in the context of an online Nuclear Masterclass. Additional material will be made available at our website\footnote{\url{https://www.york.ac.uk/physics/public-and-schools/secondary/binding-blocks/}}.


\ack
The \emph{Binding Blocks} project has been funded by EPSRC and University of York through an EPSRC Impact Acceleration Award, and by an STFC Public Engagement Small Award ST/N005694/1 and ST/P006213/1. The work of ARH is funded through grant ST/P005314/1.

%
%
%
%
%
%
%
%

\section*{References}
\bibliographystyle{iopart-num}
\bibliography{biblio}

\end{document}